\documentclass[12pt]{article}
\usepackage[dvips]{color}
\usepackage{graphicx}
\usepackage{amsthm,amssymb,amsmath,graphics,times}

\begin{document}

%\centerline{\Large \sffamily Why does it take 2 sexes to
%reproduce?}

\centerline{\Large \sffamily The Origin of 2 Sexes Through
Optimization of}

\centerline{\Large \sffamily  Recombination Entropy Against Time and
Energy}

\bigskip
\centerline{\sffamily Bo Deng\footnote{Department of Mathematics,
University of Nebraska-Lincoln, Lincoln, NE 68588, Email: {\tt
bdeng1@math.unl.edu}}}
%\centerline{\sffamily Bo
%Deng$^{\rm{1}}$, Irakli Loladze\footnote{Department of Mathematics,
%University of Nebraska-Lincoln, Lincoln, NE 68588}}

\bigskip
\noindent {\bf Abstract: Sexual reproduction in Nature requires two
sexes, which raises the question why the reproductive scheme did not
evolve to have three or more sexes. Here we construct a constrained
optimization model based on the communication theory to analyze
trade-offs among reproductive schemes with arbitrary number of
sexes. More sexes on one hand lead to higher reproductive diversity,
but on the other hand incur greater cost in time and energy for
reproductive success. Our model shows that the two-sexes
reproduction scheme maximizes the recombination entropy-to-cost
ratio, and hence is the optimal solution to the problem.}


\bigskip
\noindent{\bf Introduction.} DNA replication is a stochastic process
by which genomes mutate over time. The planetary environment of
Earth is also a dynamical process which harbors life. Since these
dynamical processes are otherwise uncoordinated, the time scale
alignment between them are important --- DNA replication must
operate at a faster time scale than Earth's environmental changes do
so that life can establish itself in a seeming constant environment,
relatively speaking, or more accurately, in punctuated equilibrium
environments (\cite{Goul77}). As a result, organisms will accumulate
too many unusable mutations during such punctuated equilibrium
states to keep their replication machinery running indefinitely.
Reproduction comes as a logical and, apparently, a practical
solution to this necessary problem of replication --- leaving a
working copy behind to continue the DNA replication process.

Reproduction did not stop at cloning. There is a net information
gain to reproduce sexually. This observation is self-evident when
examining the difference between somatic cell division (mitosis) and
reproductive cell division (meiosis). The critical difference is the
crossover process of homologous chromosomes employed by the latter.
As a result, each gamete (sperm and ovum) acquires one set of
haploid chromosomes consisting of exchanged gene alleles or DNA
segments, gene or otherwise, from both parents. Instead of one
working copy, sexually reproductive species give their offspring a
combination of two working copies of genomes --- enhancing the
genetic diversity of individuals within species. At the organismic
level however, there is a variety of costs to sexual reproduction.
The immediate ones are in time and energy. The purpose of this paper
is to quantify in what sense the sexual reproduction strategy is
better and why the logic of having a greater reproductive diversity
does not extend to 3 or more sexes.

\bigskip
\noindent{\bf Mathematical Model.} It is assumed in this paper
that DNA recombination is the principle payoff of sexual
reproduction and that information entropy at each segment of
chromosomal exchange is the measurement to quantify the payoff.
The biological importance of using entropy lies in the fact that
it measures recombinatorial diversity at the molecular level. The
scientific importance of using entropy lies in the fact it is
observer and sampling time independent, a property necessary for
being a physical law.

Recombinatorial entropy increases with the number of sexes, but it
also incurs greater cost in time and energy for reproductive
success. Our model is to find the number of sexes so that the
entropy payoff per each unit of time and energy cost is maximal, or
equivalently, the time and energy cost for each bit of
recombinatorial entropy is minimal. In other words, it is to show
that the 2-sexes reproduction scheme is the optimal solution to a
constrained optimization problem. We begin with the constraints in
terms of the following hypotheses, characterizing the crossover
process.
\begin{quote}
{\bf Recombination Model:}
\begin{enumerate}
\item There are $n$ sexes and gametes (reproductive cells) from all
$n$ sexes are required to produce a zygote. Gametes contain haploid
chromosomes and zygotes contains polyploid chromosomes, one set of
haploid chromosomes from each parental sex. %
\item Each gamete autosome (non-sex-determining chromosome)
is a mixture of $n$ parental homologous chromosomes (analogous to
the crossover
process in the case of 2 sexes).%
\item The mixing probability at any exchanging
site along any gamete autosome is the same for all parental sexes,
i.e., the equiprobability $1/n$ from each parent's contribution to
the mixing. %
\item The sex ratio of any pair of sexes is 1:1. %
\item The time and energy required to produce a zygote
is proportional to the average number of randomly grouping $n$
individuals that has exactly one sex each, called a {\em
reproductive grouping} below. In other words, the time and energy
cost for one reproductive success is inversely proportional to the
probability that a random group of $n$ individuals would be a
reproductive grouping.
\end{enumerate}
\end{quote}

Hypotheses 1 and 2 are true for $n=2$ as mentioned in the
Introduction. As for Hypothesis 3, an exchanging segment can be a
sequence of many bases or genes. The analysis below applies to
whatever length a segment may actually be. For this reason, an
exchange site can be taken as a single nucleotide base for
definitiveness throughout the discussion. The equiprobability part
of Hypothesis 3 follows from the following facts. First, when a pair
of mixed homologous chromosomes split, at any mixing segment one
copy is from one sex and the other copy is from the opposite sex.
Thus, there is always an equal number of exchanged copies from all
sexes at any site and in any population of gametes. Second, the
chromosomal crossover is independent from segment to segment so that
each gamete contains a unique mix of its contributing sex's parental
DNA. Further factoring the fact that it usually takes an
overwhelming number of gametes for each fertilization, we can indeed
assume the mixing to be completely thorough and thus the
equiprobability. As a consequence, the information entropy
(\cite{Shan48}) of the chromosomal mixing is maximal, denoted by
$H_n=\log_2 n$ in bits per segment. It can be considered to quantify
the {\em per-site crossover diversity}, referred to as the {\em
recombination entropy} below. The quantification applies to every
crossover site of all gamete autosomes. As expected, the more
parental sexes there are, the greater the recombinatorial diversity
is. As a result of this hypothesis, the model does not discriminate
against any sex's genetic contribution to reproduction.

Hypothesis 4 can be considered to be the 0th order approximation (in
the sense of \cite{Shan48}) to the sex ratio. In fact, for $n=2$ it
is a structural consequence to the fact that the sex-determining
chromosomes, $X$ and $Y$, are equally distributed in male gametes.
It is not hard to concoct hypothetical schemes to maintain the
equiratio for $n\ge 3$ cases. It should be noted that the equiratio
condition is used below to calculate the reproductive cost in a
definitive way. Changing the ratio alters the cost function from
Hypothesis 5, that in turn results in alternative models which will
not be considered further.

Hypothesis 5 should be treated as a possible scenario at an early
evolutionary stage of sexual reproduction when the sexual identity
of individual organisms was about to be well-defined and the main
cost for reproduction was to get together a reproductive group
mostly by chance. It can also be treated as a possible scenario at a
later stage of the evolution when well-defined sexual
characteristics cut down the chance encounter factor of the cost
(e.g. opposite sexes attract) which on the other hand is off-set by
like-sex interactions such as competition for mates and cooperation
for offspring rearing. Nevertheless, the hypothesis should be
treated as a ``0th order'' approximation of the cost. Cost reduction
and cost overrun mechanisms can be treated as higher order
corrections to the 0th order approximation. An analysis on the
robustness of the model against corrections is given shortly. Notice
that except for Hypothesis 5 all hypotheses are based on empirical,
textbook facts for $n=2$ (\cite{Beck02}).

The optimization objective is to maximize the recombination entropy
$H_n$ constrained to each unit of reproductive cost in time or/and
energy. When combined the problem is to maximize the dimensionless
{\em recombination entropy-to-cost} ratio $S_n=H_n/E_n$ over the
number $n$ of possible sexes, where $E_n$ is the cost in
dimensionless form (without the proportionality from Hypothesis 5).
Equivalently, the problem is to minimize the time/energy cost for
each unit of recombination entropy, $1/S_n=E_n/H_n$.

%%%%%%%%%%%%%%%%%%%%%%%%%%%%%%%%%%%%%%%%%%%%%%%%%%%%%%%%%%%%%
%%%%%%%%%%%%%%%%%%%%%%%%%%%%%%%%%%%%%%%%%%%%%%%%%%%%%%%%%%%%%
%%%%%%%%%%%%%%%%%%%%%%%%%%%%%%%%%%%%%%%%%%%%%%%%%%%%%%%%%%%%%
\begin{figure}%[ht]
\leftskip -.15in
\parbox[b]{8cm}
\noindent
{{\includegraphics[width=9cm,height=7cm]{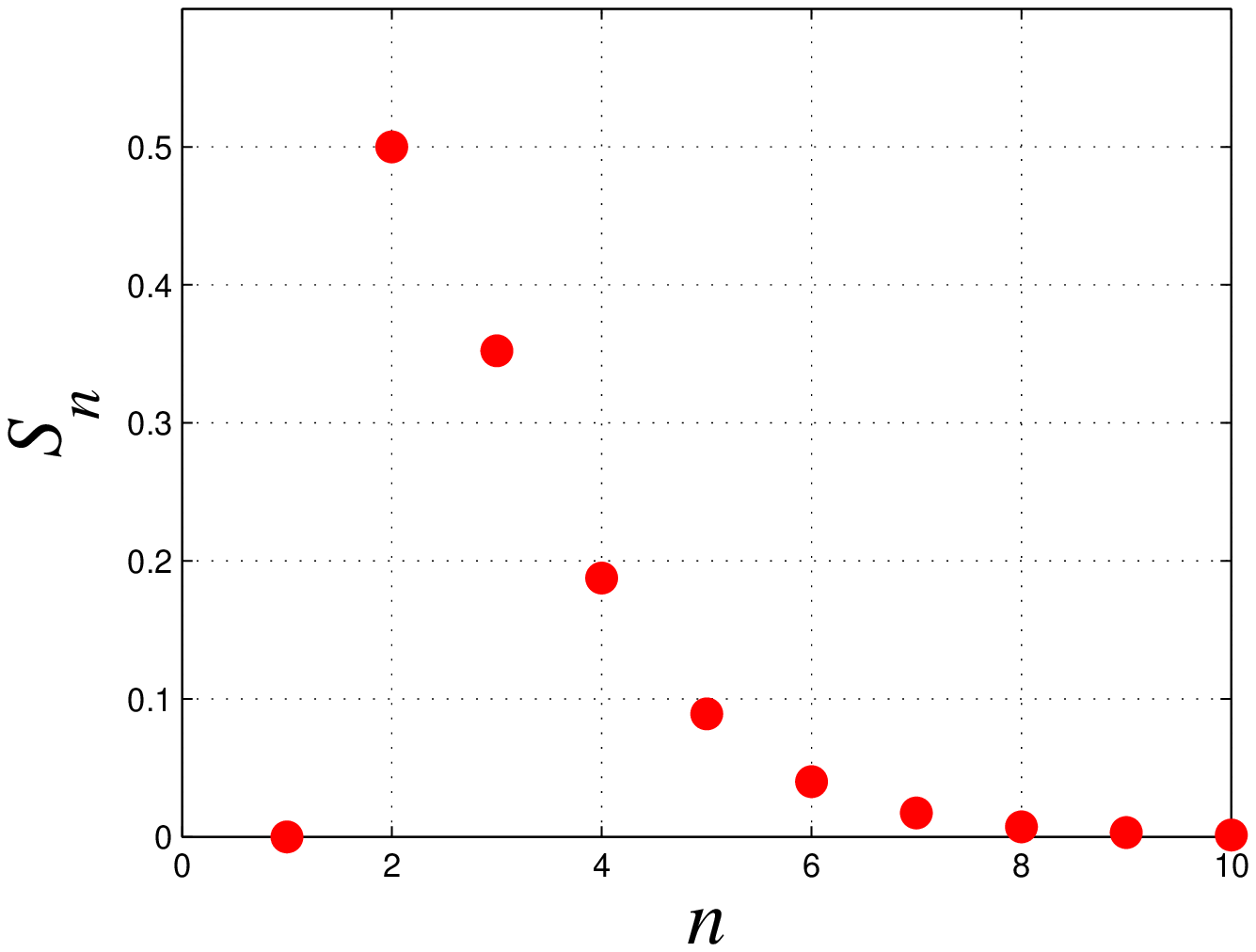}}} \hskip
.5cm
\parbox[b]{8cm}
{\centerline{\includegraphics[width=9cm,height=7cm]{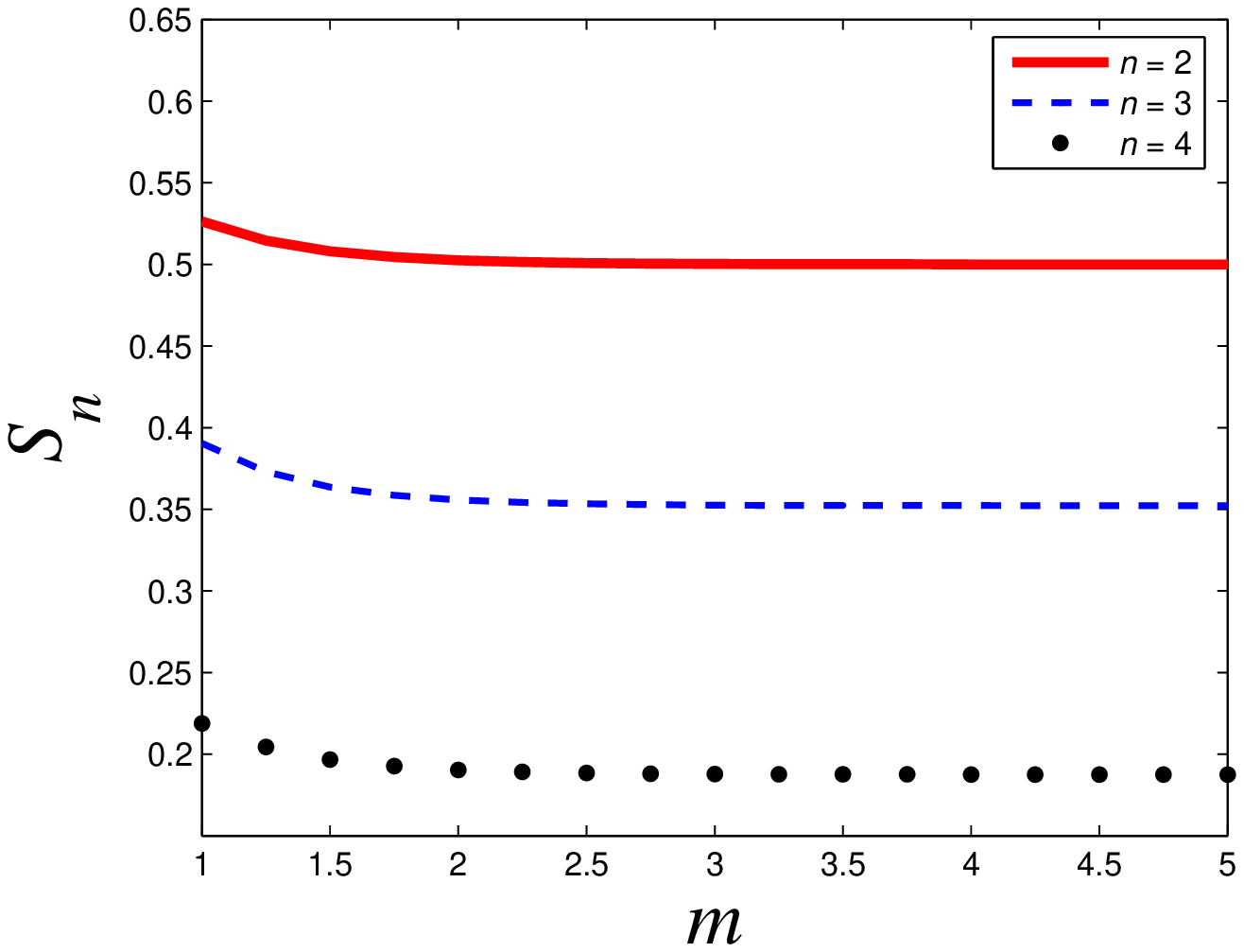}}}

\centerline{\hskip .5in (a)\hskip 3.5in(b)}

\caption{\leftskip 1in \rightskip 1in (a) Infinite population. (b)
Finite population with equal sex population $M=10^m$. \rightskip
1in }\label{fig1}
\end{figure}
%%%%%%%%%%%%%%%%%%%%%%%%%%%%%%%%%%%%%%%%%%%%%%%%%%%%%%%%%%%%%
%%%%%%%%%%%%%%%%%%%%%%%%%%%%%%%%%%%%%%%%%%%%%%%%%%%%%%%%%%%%%
%%%%%%%%%%%%%%%%%%%%%%%%%%%%%%%%%%%%%%%%%%%%%%%%%%%%%%%%%%%%%

We already have $H_n=\log_2 n$ as mentioned above. To derive $E_n$,
we proceed as follows. Without loss of generality from Hypothesis 4,
assume each sex has the same number, $M$, of individuals with $M$
being any large integer. Then there are $\left( ^{nM}_{\
n}\right)=\frac{(nM)!}{n!(nM-n)!}$ many ways to choose a group of
$n$ individuals from the total $nM$ many individuals of all sexes.
Of which only $M^n$ many are reproductive groupings by Hypothesis 1.
Hence, for each reproductive grouping there are on average ${\left(
^{nM}_{\ n}\right)}/{M^n}$ many random ways to have a group of $n$
individuals.

Throughout the discussion, all cost functions are for large
population size when $M\sim\infty$ unless stated otherwise. So
\[
E_n=\lim_{M\to\infty}\frac{\left( ^{nM}_{\ n}\right)}{M^n}.
\]
To simplify, we have
\begin{equation*}
\begin{split}
E_n &=\lim_{M\to\infty}\frac{\left( ^{nM}_{\ n}\right)}{M^n} =
\lim_{M\to\infty}{\frac{nM(nM-1)\cdots(nM-n+1)}{n!M^n}} \\
&=\lim_{M\to\infty}\frac{(n-\frac{1}{M})(n-\frac{2}{M})
\cdots(n-\frac{n-1}{M})}{(n-1)!}=\frac{n^{n-1}}{(n-1)!}.
\end{split}
\end{equation*}
Note that the time and energy cost of producing a zygote is $kE_n$
for some proportionality $k$ by Hypothesis 5. Also, $1/E_n$ is the
probability that a random group of $n$ individuals would be a
reproductive grouping. For $n=1$, $E_1=1$ as it should be for
asexual reproductive cost. For $n=2$, $E_2=2$. That is, for each
reproduction from two opposite sexes, there is one non-reproductive
interaction between like sexes. Like-sex interactions can be in the
forms of competition for mate or cooperation for offspring rearing
or just plain random encountering. Thus $E_n$ is a reasonable
functional form for reproductive cost at the population level.
Similar interpretation applies to $n>2$ cases.

As a result, the recombination entropy-to-cost ratio is
\[
S_n=\frac{H_n}{E_n}=\frac{(n-1)!\ {\rm log}_2n}{n^{n-1}}=0,\ 0.5,\
0.3522,\ 0.1875,\ 0.0892,\ 0.0399,\ 0.0072,
\]
for $1\le n\le 7$ respectively. Fig.\ref{fig1}(a) shows the graph of
$S_n$ (with $M=\infty$). Clearly, $S_2$ is the maximal solution.
That $S_1=0$ is expected since asexual reproduction has zero
recombination entropy. Fig.\ref{fig1}(b) shows the graph of $S_n =
{M^n{\rm log}_2n}/{\left( ^{nM}_{\ n}\right)}$ as a function of a
finite equal sex population $M=10^m$. The limiting ratios are good
approximations beyond a modest size $M=100$. Surprisingly, the
2-sexes reproductive strategy remains optimal even when the
population size is small, $M=10$. Notice also that the optimal
solution $S_2$ is quite robust against the next best solution $S_3$.
In fact, the difference between $S_2$ and $S_3$ is about 30\% and
43\% against $S_2, S_3$ respectively. It implies that the model can
tolerate high order corrections of considerable magnitude,
especially to Hypothesis 5, and still keep $S_2$ as the optimal
solution.

\bigskip
\noindent{\bf Discussion.} The prediction that $S_2$ is the optimal
solution is expected from any reasonable model. Some immediate
implications of our model are nevertheless surprising. With the
caveat that our Recombination Model is only a 0th order
approximation, we have the following extrapolations. Since
$S_2/S_3\sim 1.43$, a 3-sexes reproductive strategy will reduce the
per-exchange-site diversity that is due to sexual reproduction by
43\% at every evolutionary stage. Equivalently, since the reciprocal
$1/S_n$ measures the minimal time or/and energy required for each
bit of sexual recombination entropy, a 3-sexes strategy will set
back the evolutionary clock that is due to sexual reproduction by
1.5 billions years assuming life started 3.5 billions years ago
(\cite{Scho02}). All these are good reasons why a pure 3-sexes
reproductive machinery (characterized by the crossover process of
meiosis) has not been found in Nature.

There are two ways to use the Recombination Model to compare and
contrast the asexual and 2-sexes reproductive strategies since
Hypotheses 1--5 can be thought either to apply trivially to the
asexual case or not at all. In the first case, take for an example
the case of multiparous mammals which could have their litters
effortlessly cloned from one fertilized egg but did not. For them,
each gamete's recombination entropy-to-cost remains at $S_2=0.5$ in
bits per segment per cost {\em v.s.} $S_1=0$ for the would-be cloned
embryos. Since $1/S_n$ measures the minimal time or/and energy
required at the organismic level for each bit of sexual reproductive
diversity, that $1/S_1=\infty$ implies that such species, all
mammals included, would never appeared if they adopted the asexual
reproductive strategy. In this regard, our model is consistent with
this known sexual reproductive reality.

In the second case, $S_n$ cannot be used to quantify differences
between the asexual and 2-sexes reproductive strategies because
there is no recombinatorial entropy to begin with for the former.
Therefore, they must be treated as two distinct categories second
only to the primary purpose of DNA replication. Nevertheless, our
model can offer one insight into the asexual reproductive strategy.
Its continued usage can be explained by the principle reason that
sexual reproduction is not a necessary but only a sufficient way to
increase genetic diversities. Less complex organisms, such as some
bacteria which under certain conditions can speed up their mutation
rate, may be able to generate enough genomic diversity by DNA
replication alone to compensate their lack of recombinatorial
diversity. In this regard, asexual reproductive realities do not
contradict our model.

This paper takes the view that evolution is a process of time
percolation. Spacial percolation follows the least resistive passage
through a porous space. Time percolation does the same except for
the aspect that there are no predetermined or known optimal passages
in time evolution. Biological processes appeared for their own right
as optimal solutions to their immediate constraints that came before
them. It is because of this modular view on evolution that leads one
to believe that life can be stripped down to its bare minimum, say
to the ``RNA world'' in which single-strand RNA replicates itself
without any of the functions that evolved later. It is also because
of this idea of branched optimal modularization about evolution that
constrained optimization models for biological systems can be
constructed one piece a time. To demonstrate this methodology and to
stay close to the main subject of this paper, we consider two more
phenomenons of reproduction.

Although a pure 3-sexes reproduction scheme (characterized by the
crossover of three parental chromosomes) does not exist in Nature,
there are indications that sexual reproduction is not purely
2-sexual either. Under certain conditions some bacteria can freely
exchange genetic materials. It is a plausible theory that bacterial
life form was an earlier branch of all life forms on Earth (with a
possible exception of viruses) and its genetic exchange mechanism
was the precursor to the sexual reproduction scheme. Such schemes
may be modeled by the following model.
\begin{quote}
{\bf Exchange Model:}
\begin{enumerate}
\item[(a)] Any two individual organisms can exchange genetic materials.
\item[(b)] The information gain to exchange one nucleotide base among $n$
individuals is $\log_2 n$ bits per base.
\item[(c)] The time and energy needed to exchange one nucleotide base among $n$
individuals is proportional to the maintenance or existential cost
of the $n$ individuals.
\end{enumerate}
\end{quote}
Similar to the Recombination Model, the entropy-to-cost
ratio, without the dimensional proportionality is,
\[
S_n=\frac{H_n}{E_n}=\frac{\log_2 n}{n}=0,\ 0.5,\ 0.5283,\ 0.5,\
0.4644,\ 0.4308,\ 0.4011,\ 0.3750,
\]
for $1\le n\le 8$ respectively. It shows having 3 exchanging
partners is about 6\% better than having 2 or 4 exchanging partners.

\begin{figure}%[ht]
\leftskip -.15in
\parbox[b]{8cm}
\noindent
{{\includegraphics[width=9cm,height=7cm]{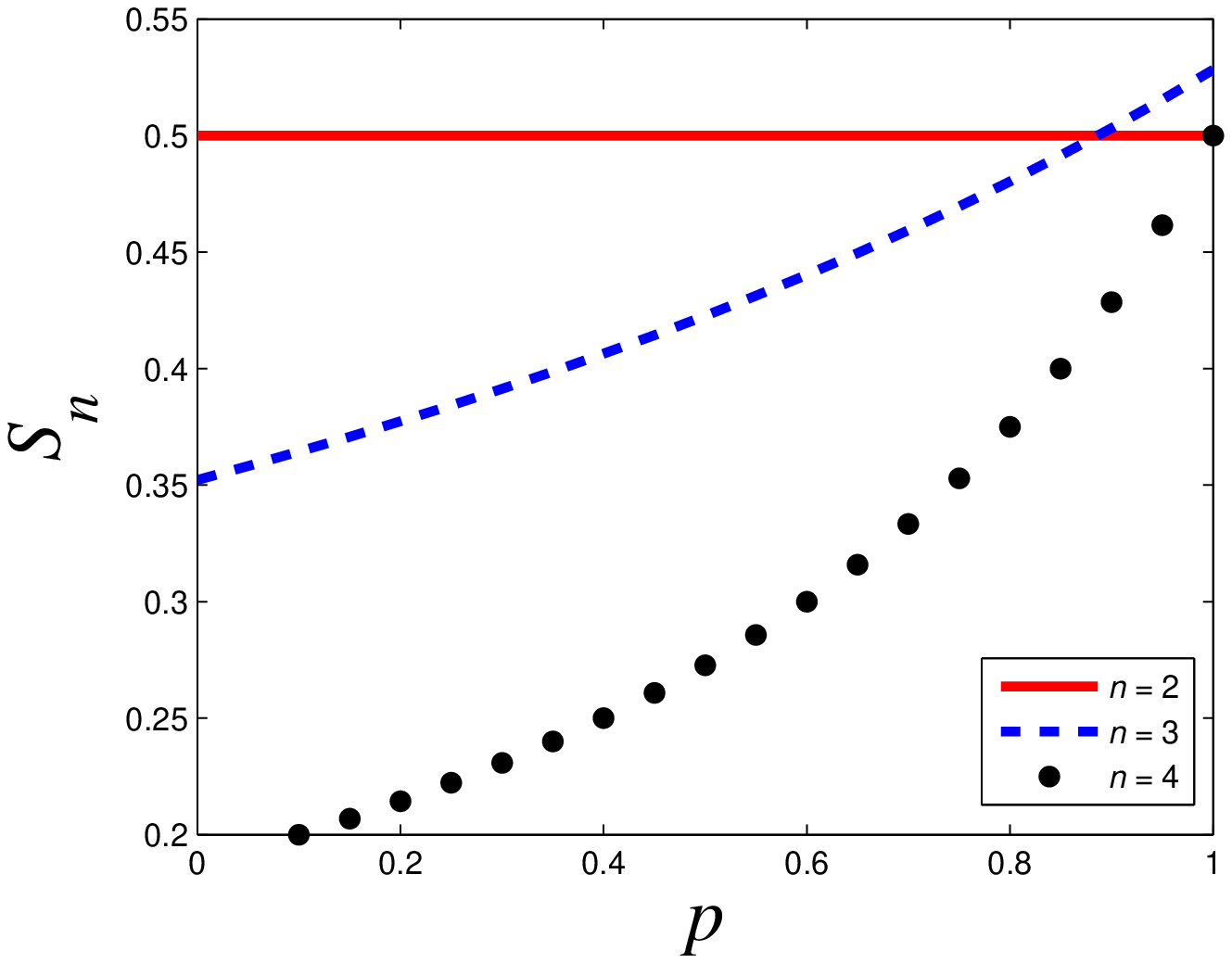}}}
\hskip .5cm
\parbox[b]{8cm}
{\centerline{\includegraphics[width=9cm,height=7cm]{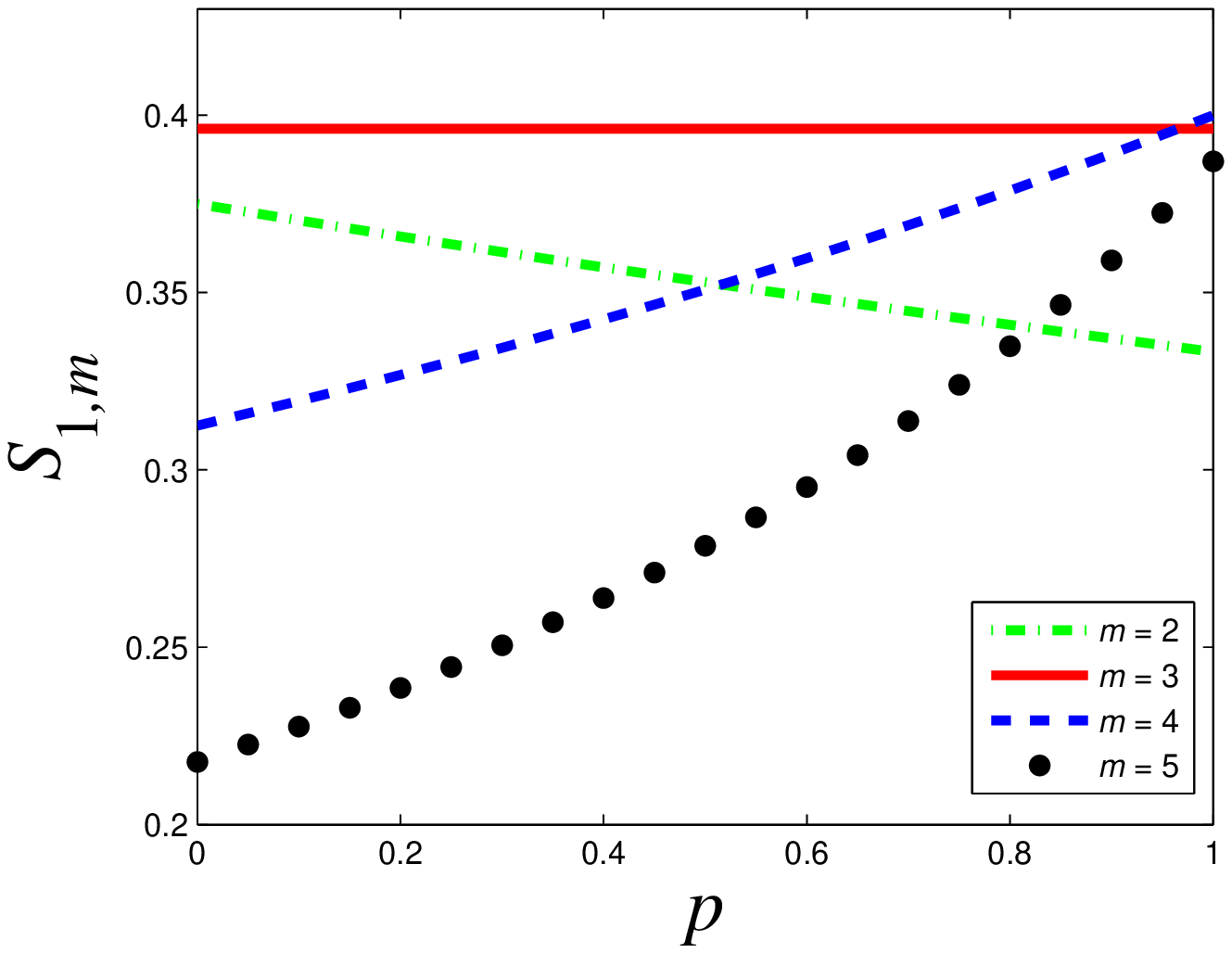}}}

\centerline{\hskip .5in (a)\hskip 3.5in(b)}

\caption{\leftskip 1in \rightskip 1in (a) The Recombination Model
with mixed cost. (b) The Multiparous Model with mixed
cost.\rightskip 1in }\label{fig2}
\end{figure}

After bacteria on the evolutionary tree came fungi. Although there
are no male or female fungi, sexually reproducing fungi reproduce by
fusing two nuclei of hyphae to produce diploid zygote which
undergoes meiosis to form haploid spores. The cost function is not
strictly that of the Exchange Model because the mating strains need
to come together before nuclei fusion can take place. It is not that
of the Recombination Model ether because of the lack of sex
distinction. The cost function can be thought as a combination of
both. More specifically, we can consider the weighted cost function
of the following form:
\[
E_n=(1-p)\frac{n^{n-1}}{(n-1)!}+pn,
\]
where $0\le p\le 1$. That $E_1=1$ holds since it must satisfy the
asexual reproduction cost condition as a default. The corresponding
reproductive entropy-to-cost ratio is
\[
S_n=\frac{H_n}{E_n}=\frac{\log_2 n}{(1-p)\frac{n^{n-1}}{(n-1)!}+pn}.
\]
Fig.\ref{fig2}(a) shows that $S_2$ is optimal for a significant
range of the weighting parameter $0\le p<0.8870$, and that $S_3$ is
optimal only when the Exchange Model is weighted heavily with
$0.8870<p\le 1$. Thus as sexual differentiation further developed so
that forming a reproductive group became critical for reproductive
success, the recombination cost must weigh more heavily, and the
Exchange-Recombination mixed model predicted the emerging of the
pure 2-sexes scheme as the optimal reproductive strategy.

%%%%%%%%%%%%%%%%%%%%%%%%%%%%%%%%%%%%%%%%%%%%%%%%%%%%%%%%%%%%%
%%%%%%%%%%%%%%%%%%%%%%%%%%%%%%%%%%%%%%%%%%%%%%%%%%%%%%%%%%%%%
%%%%%%%%%%%%%%%%%%%%%%%%%%%%%%%%%%%%%%%%%%%%%%%%%%%%%%%%%%%%%
%%%%%%%%%%%%%%%%%%%%%%%%%%%%%%%%%%%%%%%%%%%%%%%%%%%%%%%%%%%%%
%%%%%%%%%%%%%%%%%%%%%%%%%%%%%%%%%%%%%%%%%%%%%%%%%%%%%%%%%%%%%
%%%%%%%%%%%%%%%%%%%%%%%%%%%%%%%%%%%%%%%%%%%%%%%%%%%%%%%%%%%%%

Even after the establishment of the pure 2-sexes reproductive
machinery for higher organisms, reproduction is not purely
monogamous. Multiparous females mate with several males so that
members of each litter may have different fathers. The
Exchange-Recombination Model does not apply to this phenomenon
because the latter evolved well after the establishment of the
2-sexes mechanism. According to our time percolation modeling
methodology, the multiparous mechanism should be modeled based on
the {\em existence} of 2-sexes, but not entangled with the {\em
origin} of 2-sexes. Moreover, the information gain to have more male
mates for one litter is for the members of the litter at the whole
individual level while the recombination entropy gain remains the
same for each member at the molecular level. Specifically, we have
the following model:
\begin{quote}
{\bf Multiparous Model:}
\begin{enumerate}
\item[(i)] Each member of a multiparous litter is the offspring of
one of $m\ge 1$ many equally probable fathers.
\item[(ii)] The male and female sex ratio is 1:1. %
\item[(iii)] The time and energy needed to produce a zygote
is inversely proportional to the probability that a random group of
$1+m$ individuals has one female and $m$ males.
\end{enumerate}
\end{quote}
In this case, the {\em multiparous entropy} is simply
\[
H_{1,m}=\log_2 m,
\]
because of the equal probability assumption of Hypothesis (i). Thus,
with one father ($m=1$), there is no multiparous entropy gain
($H_{1,1}=0$) as expected. As for the cost, assume there are $M$
individuals of each sex for large $M\sim\infty$. There are ${\left(
^{\ 2M}_{\,m+1}\right)}$ many ways to have a group of $m+1$
individuals, of which there are $M\!\!\left( ^{M}_{\,m}\right)$ many
ways to have a group of 1 multiparous female and $m$ multiparous
males. Thus, the dimensionless cost function is approximately
\[
E_{1,m}=\lim_{M\to\infty}\frac{\left( ^{\
2M}_{\,m+1}\right)}{M\!\left(
^{M}_{\,m}\right)}=\frac{2^{m+1}}{m+1},
\]
after canceling out same factors and taking the limit in a way
similar to the derivation of the Recombination Model's cost
function. $E_{1,m}$ satisfies the pure 2-sexes condition
$E_{1,1}=E_2=2$ when there is only one father. The reproductive
entropy-to-cost ratio is
\[
S_{1,m}=\frac{(m+1)\log_2m}{2^{m+1}} = 0,\ 0.3750,\ 0.3962,\
0.3125,\ 0.2177,\ 0.1414,\ 0.0877,\ 0.0527,
\]
for $1\le m\le 8$ respectively. That is, having 3 fathers maximizes
the multiparous payoff over cost. Similar to the mixed
Exchange-Recombination cost assumption, it is reasonable to consider
the mixed multiparous cost function which is a weighted combination
of the reproductive grouping cost, $\frac{2^{m+1}}{m+1}$, and the
maintenance cost, $m+1$, both in dimensionless forms,
\[
E_{1,m}=(1-p)\frac{2^{m+1}}{m+1}+p(m+1).
\]
Fig.\ref{fig2}(b) shows that having 3 fathers is optimal for the
range of $0\le p<0.9661$ and having 4 fathers is optimal for the
rest of the weight parameter.

With regard to the concept of biological diversity, we take the view
that the grand scope of it is the sum of many constituent parts,
e.g., the constituent diversity from the combination of nucleotides
for DNA, that from the 20 amino acids for proteins, that from the
recombination of chromosomes for reproduction, and so on. The grand
term is inevitably vague, but its individual parts can be made
precise, such as the DNA replication entropy we used for the DNA
replication and the various reproductive entropies we used here for
reproduction. In addition to their definitiveness, the constituent
parts are intrinsically independent from the others in the
evolutionary time percolation sense discussed above. For example,
any reproductive entropy is delineated from the replication entropy
and the dependence of the former on the latter is only insofar as
that reproduction evolved after replication rather than the other
way around.

Arguably, in additional to these constituent diversities mentioned,
there are many more mechanistic sources contributing to biological
diversity, such as gene splicing, post-transcriptional editing,
post-translational editing, small RNA interference and regulation,
retroviral reverse transcription, heritable regulators of expression
such as methylation and acetylation, and the rate of mutations, etc.
Some of which may or may not be constant in time. Nevertheless, the
approach adopted in this paper can conceivably be applied to
construct compartmentalized models for these functions and
processes. Each can be modeled alone or in combinations with others.
To do so, one presumably has to consider some mean averages to
quantify the information exchanged within or gained by the
processes, or even to define some time dependent diversity
measurements. In such models, aggregating varying quantities or over
different phases is probably inevitable, at least for the initial
approximations of these processes. In any case, to incorporate these
processes into new models or to expand them into larger ones, one
has to formulate their mechanics in terms of hypotheses like we did
here. Each new formulation can bring out a different payoff
measurement. If the measurement is for diversity, it will be simply
added to the grand notion of diversity, with no or little effect to
other constituent parts.

The conceptual model for our model is essentially the same as that
for DNA replication introduced in \cite{Deng051,Deng052}. We treated
the latter as a communication channel when the DNA bases, $A,\ T,\
G,\ C$, are paired one at a time with their complementary bases
along the single strands of the double helix. We treated it further
as an all-purpose channel for which the mean base distribution $1/n
= \sum_{k=1}^n p_k/n$ is used for the ensemble of all genomes, each
with a particular distribution $p$ in $n$ bases. As the 0th order
approximation (\cite{Shan48}), the equiprobability gives rise to the
maximal per-base diversity in entropy $\log_2 n$, exactly the same
quantity as the per-exchange-site reproduction entropy $H_n$. We
then showed that the mean transmission/replication rate $R_n=\log_2
n/T_n$, with $T_n$ the mean base pairing time, is maximal for $n=4$
provided that the pairing time of the hydrogen bonds of the $GC$
pair is between 1.65 and 3 times that of the $AT$ pair. Although our
reproduction models here do not mechanistically fit to a
communication model as neatly as the replication model does, the
apparent analogy is hard to miss. For example, each reproduction
from the Recombination Model can be thought as one packet
transmission, and each packet contains a total information
$H_n\times L$, where $L$ is the number of mixing sites along gamete
chromosomes. Since it is more natural to consider the whole
reproduction cost $E_n$ in time or/and energy rather than to unitize
it, $E_n/L$, we are led to the normalized recombination
entropy-to-cost ratio $S_n=H_n/E_n$, which is now clearly
proportional to the mean transmission rate $R_n$ if we think the
reproduction as a channel and $E_n$ as the time.

Based on the game theory (\cite{Owen68,Bomz89}), the idea of
Evolutionarily Stable Strategy (ESS, \cite{Smit82}) was also used to
explain sex evolution at the phenotypic level, in particular the
problem of 1:1 sex ratio. Our method is based on Shannon's
communication theory (\cite{Shan48}). However, both methods are
rooted in constrained optimizations. The game theory is about
maximizing fitness payoffs with play rule constraints. The
communication theory is about maximizing information with
constraints in transmission time or storage space or energy
consumption. Our view is that ESS is a more plausible theory for
species behaviorial interaction at the community level. In contrast,
the 1:1 sex ratio strategy most sexually reproducing organisms adopt
has a more fundamental root at the molecular level for reproduction.
Otherwise, ESS would have to predict unequal sex ratio for species
such as African lions and elephants for which a large portion of
males do not procreate.

The stochastic formulation of both replication and reproduction
models implies the following. (1) Since parents cannot choose the
genetic composition of their offspring (Hypothesis 3) and offspring
cannot choose its parents (Hypothesis 5), the notion of ``individual
ownership of DNA'' cannot be well-defined. Thus each organism is
only an accidental and temporary carrier of the protogenic DNA at
the origin or origins of life, echoing with Darwin's ``common
descent'' theory (\cite{Darw59}). (2) The maximal recombination
entropy in bits per base or segment resides in every organism in a
suspended probabilistic state and at every moment of recombination
in time, in the averaged sense of the entropy definition. It is by
replication and reproduction that the maximal entropies are
expressed through time, expanded in length, and multiplied in space.
(3) Because the 4-base replication strategy and the 2-sexes
reproduction strategy are optimal strategies, evolution is where it
should be in time as far as the part of the biological diversity due
to replication and reproduction is concerned, though individual
organisms and species are accidental. (4) Since Hypothesis 5 leads
to the notion of minimizing reproductive cost, our model is
consistent with the observation that sex-specific features have the
effect of cost reduction from same sex interactions.

\bigskip

\noindent {\bf Acknowledgement:} The author gratefully acknowledges
comments and suggestions from Irakli Loladze, Lawrence G. Harshman
of UNL. Special thanks go to the referees whose comments and
suggestions led to the inclusion of the Exchange Model and the
Multiparous Model.

\end{document}